\documentclass[aps,prl,twocolumn,reprint,floatfix,groupedaddress]{revtex4}
\usepackage[english]{babel}
\usepackage{amsmath, amsthm, amssymb, latexsym, graphicx}
\usepackage[utf8]{inputenc}
\usepackage{hyperref}

\usepackage{color}

\def\bb{{\bf B}}
\def\vv{{\bf v}}
\def\vA{{\bf v_A}}

\def\xx{{\bf x}}
\def\zz{{\bf z^\pm}}
\def\zzp{{\bf z^+}}

\def\zzm{{\bf z^-}}
\def\be{\begin{equation}}
\def\ee{\end{equation}}
\def\ba{\begin{eqnarray}}
\def\ea{\end{eqnarray}}

\def \pmbtext#1{\leavevmode
     \setbox0\hbox{#1}
     \kern0,4pt \copy0 \kern-\wd0
     \kern-0,2pt \raise0,3pt \box0 }

\begin{document}

\title{Compressible magnetohydrodynamic turbulence in the Earth's magnetosheath: estimation of the energy cascade rate using {\it in situ} spacecraft data}

\author{L. Z. Hadid}
\affiliation{Swedish Institute of Space Physics, SE 751 21, Uppsala, Sweden}
\affiliation{LPP, CNRS, Ecole Polytechnique, Univ. UPMC, Univ. Paris-Sud, Observatoire de Paris, Universit\'e Paris-Saclay, Sorbonne Universit\'e, PSL Research University, 91128 Palaiseau, France}
\author{F. Sahraoui}
\affiliation{LPP, CNRS, Ecole Polytechnique, Univ. UPMC, Univ. Paris-Sud, Observatoire de Paris, Universit\'e Paris-Saclay, Sorbonne Universit\'e, PSL Research University, 91128 Palaiseau, France}
\author{S. Galtier}
\affiliation{LPP, CNRS, Ecole Polytechnique, Univ. UPMC, Univ. Paris-Sud, Observatoire de Paris, Universit\'e Paris-Saclay, Sorbonne Universit\'e, PSL Research University, 91128 Palaiseau, France}
\affiliation{Departement of Physics, University of Paris-Sud, Orsay, France}
\author{S. Y. Huang}
\affiliation{School of Electronic Information, Wuhan University, Wuhan, China}
\date{\today}

\begin{abstract}
The first estimation of the energy cascade rate $\vert\epsilon_C\vert$ of magnetosheath turbulence is obtained using the CLUSTER and THEMIS spacecraft data and an exact law of compressible isothermal magnetohydrodynamics turbulence. $\vert\epsilon_C\vert$ is found to be of the order of $ 10^{-13} J.m^{-3}.s^{-1} $, at least two orders of magnitude larger than its value in the solar wind (order of $10^{-16} J.m^{-3}.s^{-1} $ in the fast wind). Two types of turbulence are evidenced and shown to be dominated either by incompressible Alfv\'enic or compressible magnetosonic-like fluctuations. Density fluctuations are shown to amplify the cascade rate and its spatial anisotropy in comparison with incompressible Alfv\'enic turbulence. Furthermore, for compressible magnetosonic fluctuations, large cascade rates are found to lie mostly near the linear kinetic instability of the mirror mode. New empirical power-laws relating $\vert\epsilon_C\vert$ to the turbulent Mach number and to the internal energy are evidenced. These new finding have potential applications in distant astrophysical plasmas that are not accessible to {\it in situ} measurements.
\end{abstract}

\maketitle

Turbulence is a ubiquitous non-linear phenomenon in hydrodynamic and plasmas flows that transfers dynamically energy between different scales. In astrophysical plasmas, turbulence is thought to play a major role in various processes such as accretion disks, star formation, acceleration of cosmic rays, solar corona and solar wind heating, and energy transport in planetary magnetospheres \cite{Zimbardo2006,Schekochihin2009,Papen2014, Tao2015}. Thanks to the availability of {\it in situ} measurements recorded on board various orbiting spacecraft, the solar wind and the Earth's magnetosheath (i.e., the region of the solar wind downstream of the bow shock) provide a unique laboratory for the observational studies of plasma turbulence. An important feature of magnetosheath turbulence is the high level of density fluctuations in it, which can reach $\sim 50\%-100\%$~\cite{Song1992,Sahraoui2003,lucek2005,Genot2009} in comparison with  $\sim 5\%-20\%$ in the solar wind~\cite{howes2012,Hadid2017}. This makes the magnetosheath a key region of the near-Earth space where significant progress can be made in understanding compressible plasma turbulence, which is poorly understood although it is thought to be important in various astrophysical plasmas, such as supernovae remnants or the interstellar medium (ISM)~\cite{Vasquez1996,LithwickGoldreich2001,Cho02,Federrath2010,Zank2017}. 

In the solar wind the magnetohydrodynamics (MHD) approximation has been successfully used to study turbulence cascade at scales larger than the ion inertial length (or Larmor radius)~\cite{TuMarsch1995,Bruno2005}. As in neutral fluid turbulence, an inertial range of MHD turbulence is generally evidenced by the observation of a power spectral density (PSD) exhibiting
a power-law over a wide range of scales. This power-law is a manifestation of a turbulent cascade of energy 
from large scales, where the energy is injected, to the smaller ones where the energy is dissipated. The energy transfer over scales is assumed to occur at a constant rate, which is equal to the rate at which energy is injected and dissipated into the system. This quantity carries therefore a major importance in modeling the processes of particle acceleration and heating in plasmas since it provides an estimation of the amount of energy that is eventually handed to the plasma particles at the dissipation scales~\cite{Sahraoui2009}. Within the incompressible MHD turbulence theory, the energy cascade rate can be estimated using the so-called third-order law relating the longitudinal structure functions of the magnetic and the velocity fields, to the spatial scale $l$~\cite{PolitanoPouquet1998} (PP98 hereafter). The PP98 model has been used to estimate the cascade rate in the solar wind from the ACE and ULYSSES spacecraft data~\cite{Smith2006d,Sorriso2007,Vasquez2007,MacBride2008,Stawarz2009,Coburn2012,Coburn2015}. Those estimations were used to better understand the long-standing problem of the non-adiabatic heating of the solar wind observed at different heliospheric distances ($\sim 0.3-100$ Astronomical Unit) by the Helios~\cite{Marsch1982,Freeman1988}, Pioneer and Voyager spacecraft~\cite{Gazis1984,Gazis1994,Richardson1995}. To date no such estimation of the cascade rate exists for magnetosheath turbulence. The main reason for that being the complex nature of magnetosheath turbulence and the importance of density fluctuations in it, which requires going beyond the PP98 model to include compressibility. Recently, an exact law of compressible isothermal MHD turbulence has been derived \cite{Banerjee2013} (hereafter BG13). It has been successfully used to improve our understanding of the role of density fluctuations in heating the fast and slow solar wind by showing in particular that, even if they are weak and represent only $\sim 5\%-20\%$ of the total fluctuations, they nevertheless can enhance significantly the turbulence cascade rate~\cite{Hadid2017,Banerjee2016}.

In the present Letter, we provide the first estimate of the energy cascade rate of compressible MHD turbulence in the Earth's magnetosheath using the BG13 and {\it in situ} wave and plasma data. We investigate furthermore how density fluctuations amplify the energy cascade rate and how they affect its spatial anisotropy. The role of density fluctuations is highlighted by comparing the results obtained from the compressible BG13 and the incompressible PP98 exact laws. Under the assumptions of time stationarity, space homogeneity and isotropic turbulence, the PP98 exact law is given by

\be
- \frac{4}{3} \varepsilon_{I} \ell = 
\left\langle {\left(\delta \zzp \right)^2 \over 2} \delta z_{\ell}^- + {\left(\delta \zzm \right)^2 \over 2} \delta z_{\ell}^+ \right\rangle \rho_0 \, , \label{pp98a}
\ee

\noindent and the BG13 model is given by
\ba
-{4 \over 3} \varepsilon_C \ell  &=&\left\langle \frac{1}{2} \left[ \delta ( \rho \mathbf{z}^-) \cdot \delta \mathbf{z}^-  \right]  {\delta  {z}_{\ell}^+}
+  \frac{1}{2} \left[  \delta ( \rho  \mathbf{z}^+) \cdot \delta \mathbf{z}^+ \right]  {\delta {z}_{\ell}^-} \right\rangle \nonumber \\
&+& \left\langle 2  \delta \rho \delta e  \delta v_{\ell} \right\rangle \nonumber \\
&+& \left\langle 2 {\overline{\delta} \left[ \left(1 + \frac{1}{\beta} \right) e + { v_A^2  \over 2}\right] \delta ( \rho_1 v_{\ell})}  \right\rangle
\label{fcphi}
\ea 
\noindent where  $\zz = \vv \pm \vA$ represent the Els\"asser variables, $\vv$ being the plasma flow velocity, $\vA \equiv \bb/\sqrt{\mu_0 \rho}$ is the Alfv\'en speed, $\rho = \rho_0 + \rho_1$ is the local plasma density ($\rho_0=\langle \rho \rangle$ and $\rho_1$ are the mean and fluctuating density), $\delta \zz \equiv \zz (\xx + \boldsymbol{\ell}) - \zz (\xx)$ is the spatial increment of $\zz$ at a scale $\ell$ in the radial direction, $\left\langle ... \right\rangle$ is the ensemble average, $\overline{\delta} \psi \equiv (\psi (\xx + \boldsymbol{\ell}) + \psi (\xx))/2$, $e=c_s^2 \ln (\rho /\rho_0)$ is the internal energy, with $c_s$ the constant isothermal sound speed,  and $\beta = 2 c_s^2 / v_A^2$ is the local ratio of the total thermal to magnetic pressure ($\beta=\beta_e+\beta_p$). Note that in the PP98 model, $\rho$ is replaced by $\rho_0$ in the definition of the Alfv\'en speed. The reduced form of BG13 used here assumes furthermore the statistical stationarity of the plasma $\beta$ and the negligible contribution of the energy source terms w.r.t. flux terms~\cite{Banerjee2016}. It is worth noting that contrary to incompressible MHD theory, the BG13 compressible model yields an energy cascade rate that is not related only to third-order moments of the different fields increments but rather involves more complex combinations of the turbulent fields. In particular the last term in  the RHS of equation \ref{fcphi}, is written as a first order increment multiplied by an averaged quantity $\overline{\delta} \psi$. This term that plays a leading order in the BG13 model~\citep{Banerjee2016,Hadid2017} is likely to converge faster than the usual third-order terms when estimated from spacecraft observations or simulations data.

{\it Results ---} The data used here were measured by the CLUSTER and THEMIS B/C spacecraft~\cite{Escoubet1997,Angelopoulos2008}. Combining the data from the two missions was aimed at increasing the sample size of our study for a better statistical convergence. The magnetic field measurements of CLUSTER come from the Flux Gate Magnetometer (FGM) \cite{Balogh1993}, while the ion and electron plasma moments (density, velocity and temperature) come from the Cluster Ion Spectrometer (CIS) experiment~\cite{REME1997}, and the Plasma Electron and Current Experiment (PEACE) \cite{Johnstone1997}, respectively. A special attention has been paid to the reliability of the plasma data, in particular the plasma density measurements, through cross-checks between the PEACE, CIS and the WHISPER experiments~\cite{Trotignon2003} on board Cluster (we selected only intervals when the instruments  were consistent with each other). For Themis spacecraft, the magnetic field data and the plasma moments are measured respectively by the FGM \cite{Auster2009} and the ElectroStatic Analyzer (ESA) \cite{McFadden2009}. In selecting our samples, we eliminated time intervals that contained significant disturbances or velocity shears and considered only time intervals that have a relatively stationary plasma $\beta$ (as discussed above)~\citep{Hadid2017}.

\begin{figure}
\includegraphics[width=1\columnwidth]{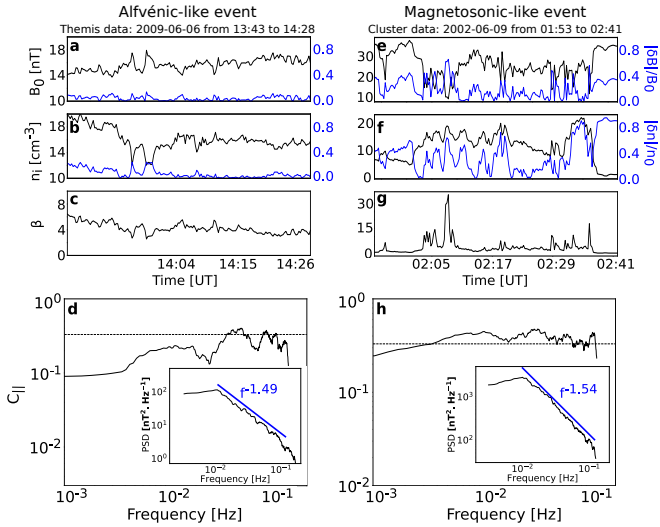}
\caption{(a, e) The magnetic field modulus (black) and fluctuations (blue), (b, f) ion number density (black) and density fluctuations (blue) and (c, g) total plasma $\beta$. (d) and (h) the corresponding magnetic compressibility. The inset is the PSD of $\delta B$ and the corresponding power-law fit in the inertial range (blue).}
\label{EF_Examples}  
\end{figure}

A large statistical survey of PSD of $\delta B$ in the magnetosheath using the Cluster data showed that only a small fraction ($17\%$) had a scaling close to the Kolmogorov spectrum $f^{-5/3}$ at the MHD scales, and were dominated either by incompressible Alfv\'enic or compressible magnetosonic fluctuations~\cite{Huang2017}. Those data sets correspond to a state of fully developed turbulence that is reached away from the bow shock toward the magnetosheath flanks. The remaining cases were found to have shallower spectra close to $f^{-1}$ and were distributed essentially near the bow shock toward the nose of the magnetopause. \citet{Hadid2015} showed that the $f^{-1}$ spectra were populated by uncorrelated fluctuations. Here, since we are interested in estimating the energy cascade rate, we focus only of the Kolmogorov-like cases that correspond to a fully developed turbulence where an inertial range can be evidenced. The final data selection resulted in $47$ time intervals of equal duration $48 \, \text{mn}$, which corresponds to a number of data points $N \sim 240$ in each interval with a $12 \, \text{s}$ time resolution ($N_{tot}\sim 2\times 10^4$). The resulting samples were divided into two groups depending on the nature of the dominant turbulent fluctuations: incompressible Alfv\'enic-like and compressible magnetosonic-like ones. This was done using the magnetic compressibility $C_{||}=\delta B_{||}^2/\delta B^2$ (i.e., the ratio between the PSDs of the parallel to the total magnetic fluctuations; parallel being along the mean background field ${\bf B}_{0}$)~\cite{Gary2009a,Salem2012,Sahraoui2012,Hadid2015}. Figure \ref{EF_Examples} shows two  examples of an Alfv\'enic-like event characterized by a nearly constant $B$, a subdominant $\delta B_{||}$ and weak density fluctuations, and a compressible case having large $B$ and density fluctuations and a strong $\delta  B_{||}$. 

\begin{figure}
\includegraphics[width=1\columnwidth]{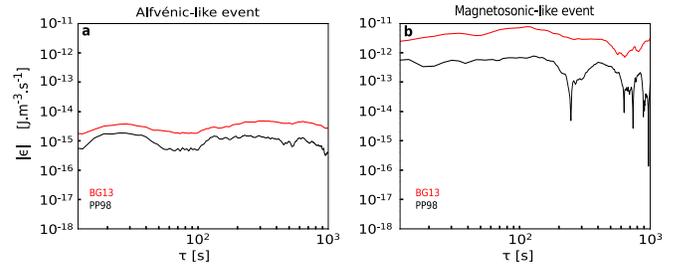}
\caption{The energy cascade rates computed using BG13 (red) and PP98 (black) for the same (a) Alfv\'enic and (b) magnetosonic-like events of Figure \ref{EF_Examples}.}
\label{EF}  
\end{figure}

\begin{figure}
\includegraphics[width=1\columnwidth]{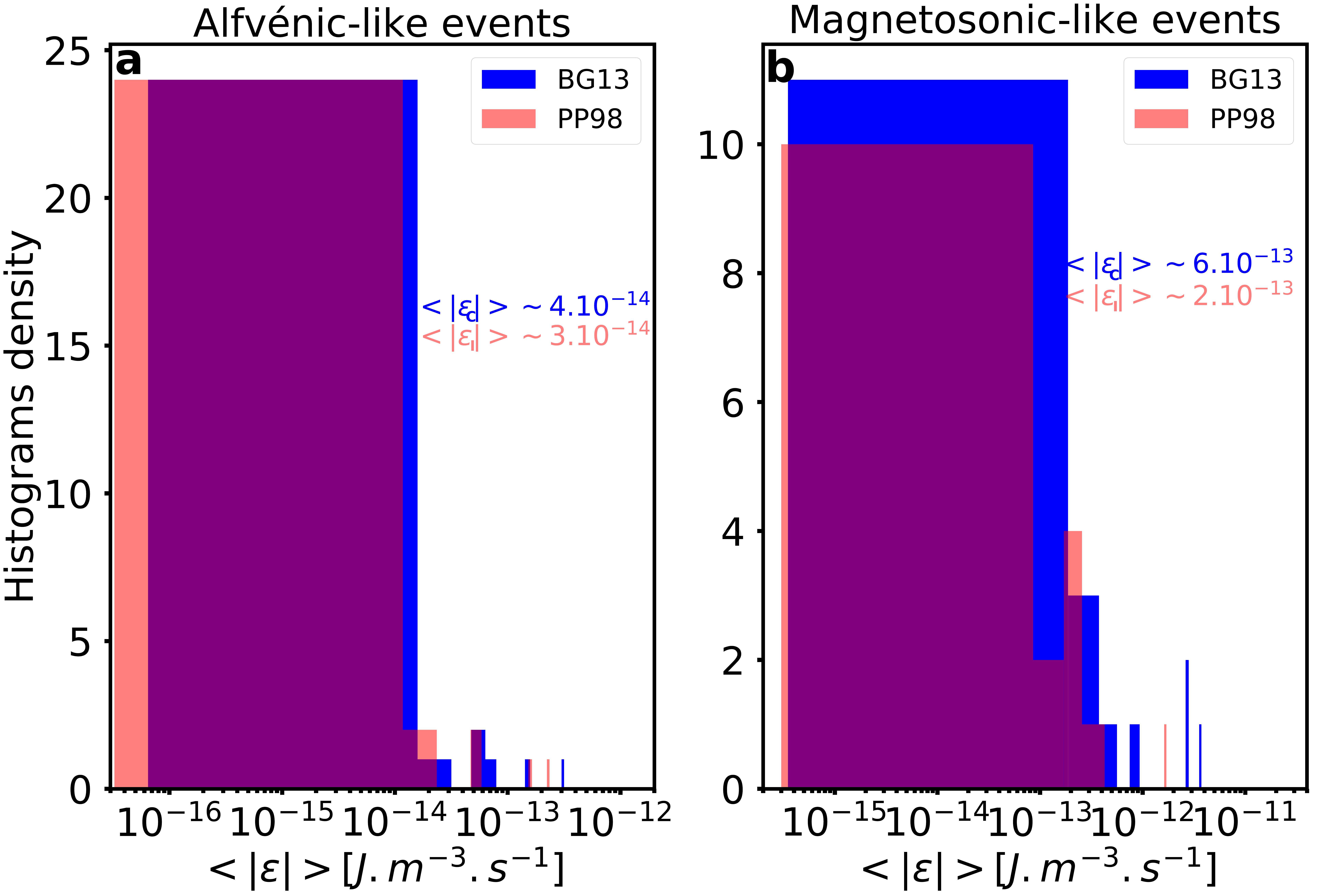}
\caption{Histograms of $\langle \vert \epsilon_{I}\vert \rangle$ (red) and $\langle \vert \epsilon_{C}\vert \rangle$ (blue) computed using the exact laws of PP98 and BG13 for the (a) Alfv\'enic and (b) magnetosonic-like events.}
\label{EF_Histo}  
\end{figure}

For each of these groups we computed the absolute values of the cascade rates $\vert \epsilon_C \vert$ and $\vert \epsilon_I \vert$ from the compressible BG13 and the incompressible model PP98, respectively. To do so, temporal structure functions of the different turbulent fields involved in equations \ref{fcphi} and \ref{pp98a} were constructed for different values of the time lag $\tau$ between $10s$ and $1000s$ in order to probe into the scales of the inertial range. In this study we considered magnitude of the cascade rate rather than its signed value (assuming that the former is statistically representative of the actual cascade rate). This is because signed cascade rates require very large statistical samples to converge~\citep{Coburn2015, Hadid2017}, which are not available to us for this study. However, by applying a linear fit on the resulting energy cascade rates, we considered only the ones that are relatively linear with $\tau$ and showed no sign change at least over one decade of scales in the inertial range. Two main observations can be made from the two examples shown in Figure~\ref{EF}: first, the incompressible cascade rate $\vert \epsilon_I \vert $ is larger by a factor $\sim 100$ in the magnetosonic case compared to the Alfv\'enic one, which can be explained by the large amplitude $\delta B$ in the former~\cite{Hadid2017}. Second, density fluctuations in the magnetosonic case amplify $\vert \epsilon_C \vert$ by a factor $\sim 7$  w.r.t. $\vert \epsilon_I \vert $. The results of analysis of all the samples are summarized in Figure~\ref{EF_Histo}. As one can see in that figure, for the incompressible Alfv\'enic cases, the histograms of $\langle \vert \epsilon_C \vert \rangle $ (blue) and $\langle \vert \epsilon_I \vert \rangle $ (red), almost overlap and the mean values for both is of the order of $\sim 10^{-14} \, J.m^{-3}.s^{-1}$, whereas for the compressible magnetosonic events the histogram of $\langle \vert \epsilon_C \vert\rangle$ (blue) is shifted towards larger values compared to $\langle \vert \epsilon_I \vert\rangle$ (red). The corresponding mean values are respectively  $\sim 6 \times 10^{-13} \, \text{and} \, \sim 2 \times 10^{-13} J.m^{-3}.s^{-1}$. We note that those values should be considered with attention, since most of the samples lie below $\langle \vert \epsilon \vert\rangle$.

\begin{figure}
\includegraphics[width=1\columnwidth]{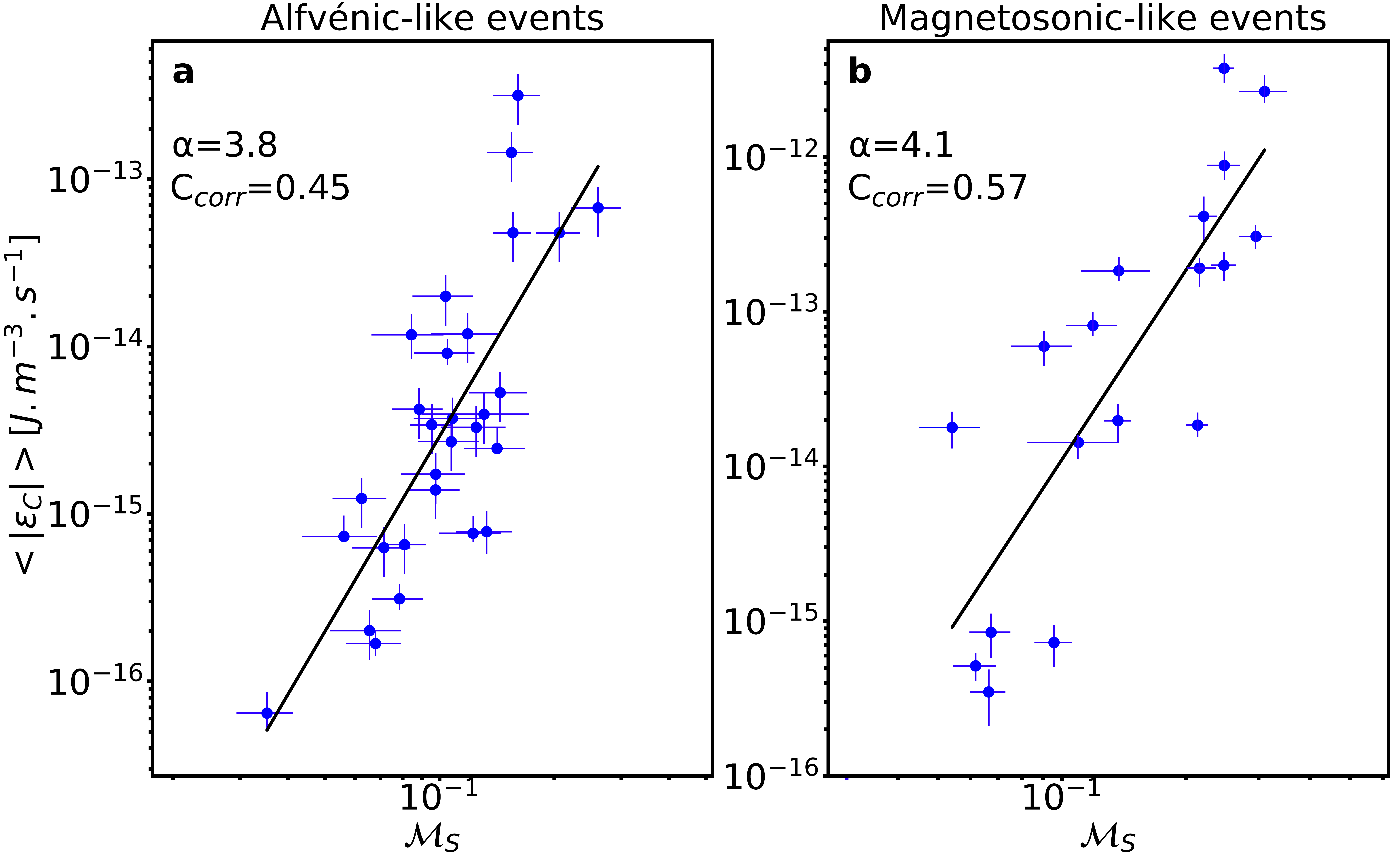}
\caption{Compressible energy cascade rate as a function of the turbulent Mach number for the Alfv\'enic (a) and Magnetosonic-like (b) events. The black line represents a least square fit of the data, $\alpha$ is the slope of the power-law fit. The error bars represent the standard deviation of $|\epsilon_{C}|$.}
\label{Mach_Number}  
\end{figure}

Interestingly, the role of the compressibility in increasing the compressible cascade rate can be evidenced by the turbulent Mach number ${\cal M}_{s}=\sqrt{<{\delta v}>^2/{c_s^2}}$, where $\delta v$ is the fluctuating flow velocity. Figure~\ref{Mach_Number} shows a power law-like dependence of $\langle \vert \epsilon_C \vert \rangle $ on ${\cal M}_{s}$ as $\left\langle\vert\varepsilon_C\vert\right\rangle \sim {\cal M}_{s}^{4}$, steeper than the one observed in the solar wind \cite{Hadid2017}. To the best of our knowledge there are no theoretical predictions that relate $\varepsilon$ to ${\cal M}_{s}$ in compressible turbulence. However, in incompressible flows, dimensional analysis {\it \`a la} Kolmogorov yields a scaling that relates $\varepsilon_I$ to the third power of ${\cal M}_{s}$. The high level of the density fluctuations in the magnetosheath seems to modify this scaling to the one we estimated here. Although more analytical and numerical studies are needed to understand the relationship between $\vert\varepsilon_C\vert$ and ${\cal M}_{s}$, the scaling law obtained here may be used as an empirical model for other compressible media non accessible to {\it in-situ} measurements.

\begin{figure}
\includegraphics[width=1\columnwidth]{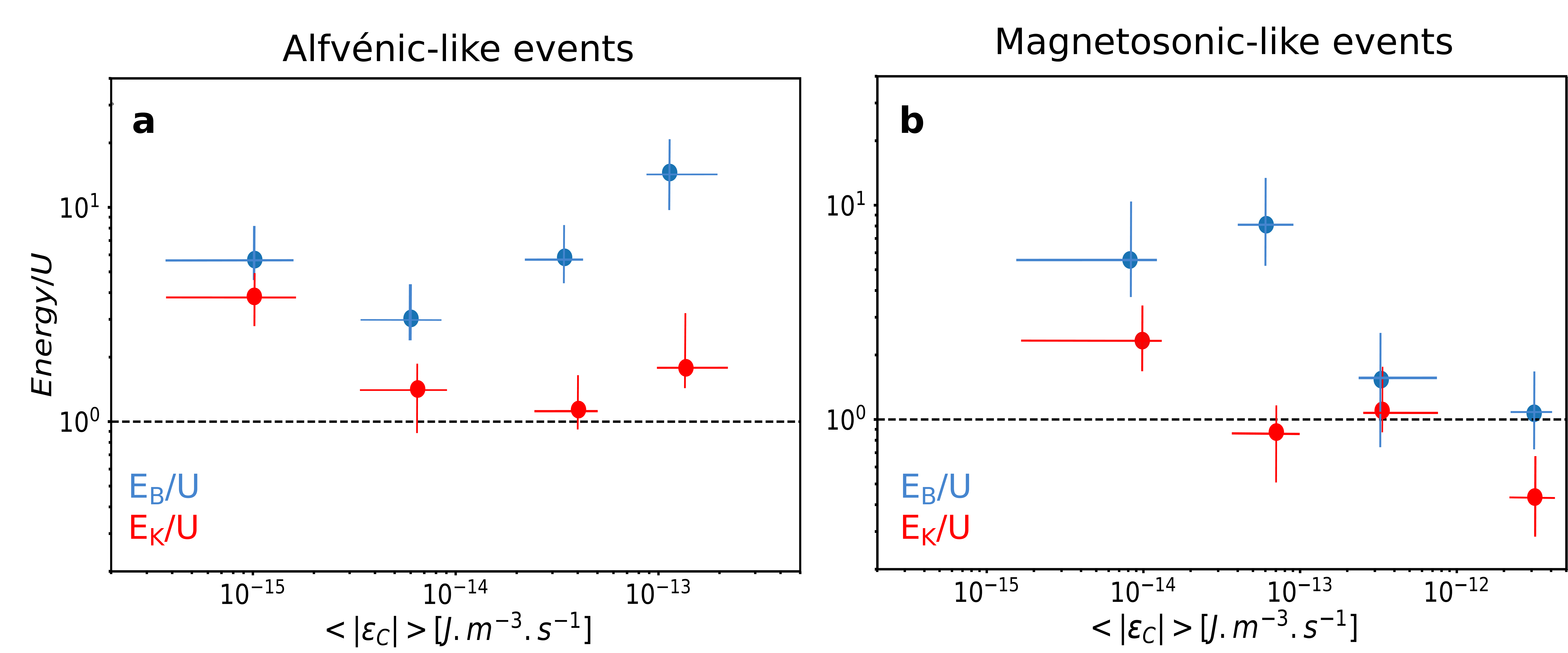}
\caption{Normalized mean compressible magnetic (blue) and kinetic (red) binned energies to the internal energy as a function of the binned compressible energy cascade rate for the Alfv\'enic (a) and magnetosonic (b) events.}
\label{Energies}  
\end{figure}

A good correlation is also evidenced between the leading order of compressible internal energy $U=\rho_0c_s^2 \ln(1+\rho_1/\rho_0$) of the turbulent fluctuations and the cascade rate in the magnetosonic-like events. Figure \ref{Energies} shows for each type of turbulent fluctuations the dependence of $\langle |\epsilon_{C}|\rangle$ on the normalized (to $U$) kinetic energy $E_K=\frac{1}{2}\rho_0 \delta v^2$ and magnetic energy $E_B=\frac{1}{2\mu_0}\delta B^2$, where $\delta B$ and $\delta v$ are respectively the fluctuating magnetic and flow velocity fields. First, one can see that for both types of turbulence, $E_B/U$ (blue) dominates over $E_K/U$ (red). Moreover, for the magnetosonic-like cases there is a general trend indicating that high $\langle |\epsilon_{C}|\rangle$ corresponds to increasing compressible internal energy, which becomes comparable (or slightly larger) than the kinetic and magnetic energies at the highest values of $\langle |\epsilon_{C}|\rangle$. This trend is not seen on the Alfv\'enic cases indicating the prominent role of the internal energy in controlling the cascade rate. This last result contrasts significantly with the finding in the solar wind~\cite{Hadid2017}.

\begin{figure}
\includegraphics[width=1\columnwidth]{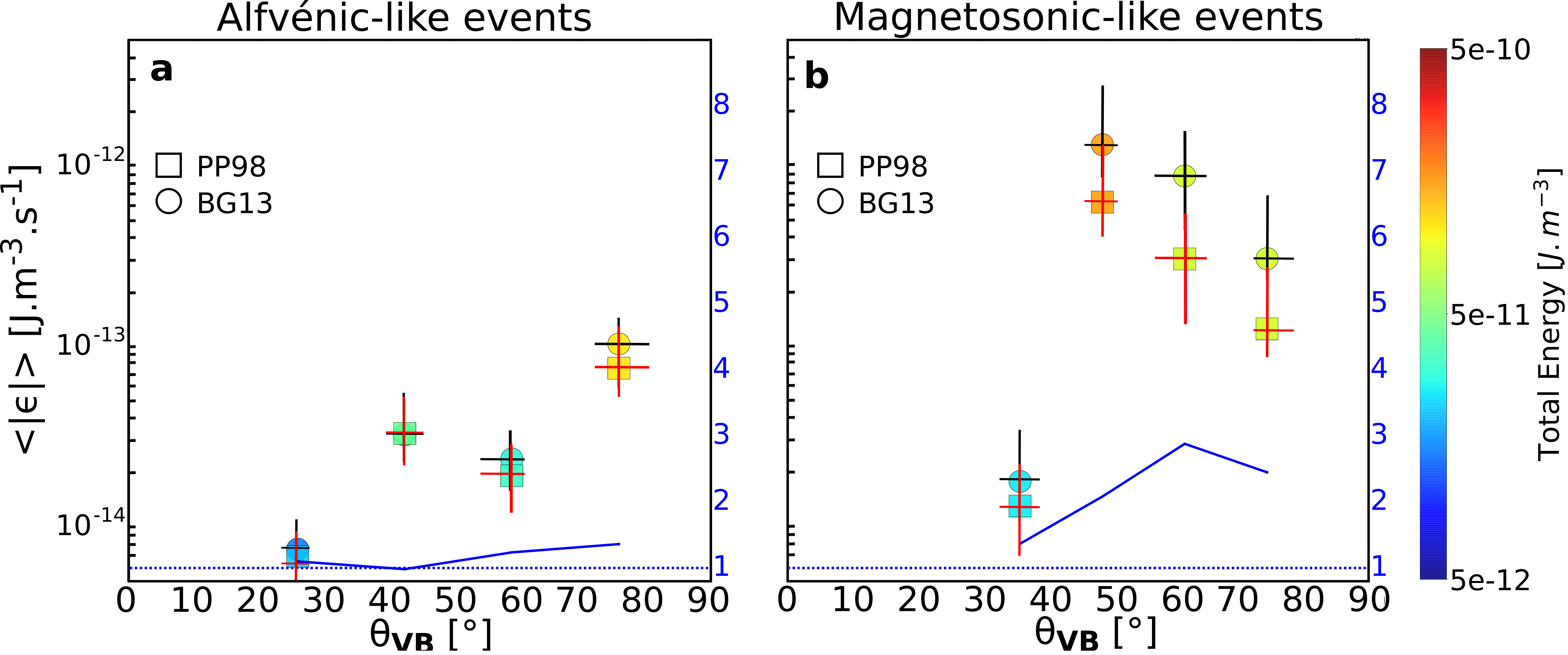}
\caption{Compressible and incompressible energy cascade rates $\langle \vert \epsilon_{C}\vert \rangle $ and  $\langle \vert \epsilon_{I}\vert \rangle $ as a function of the mean angle $\Theta_{\bf VB}$ and the total energy (colored bar) for the Alfv\'enic (a) and magnetosonic (b) events. The blue line is the ratio ${\cal R}=\langle \vert \epsilon_C \vert\rangle/ \langle \vert \epsilon_I \vert \rangle$. }
\label{Angle_Anisotropy}  
\end{figure}

To study the anisotropic nature of the cascade rate for the different types of the MHD fluctuations, we examine the dependence of the estimated cascade rates on the mean angle $\Theta_{\bf VB}$ between the local magnetic and flow vectors. This approach has been already used in similar studies of solar wind turbulence \cite{Smith2009,Marino2012,Hadid2017}. Here we consider only the events that have a relatively uniform $\Theta_{\bf VB}$ to guarantee that the spacecraft is sampling nearly the same direction of space for each time interval (using the Taylor frozen-in flow assumption). As one can see in Figure \ref{Angle_Anisotropy}, for both models the cascade rate is lower in the parallel direction than in the perpendicular one. The same trend is observed for the total energy, except that for the magnetosonic events the highest cascade rate and total energy are observed at oblique angles $\Theta_{\bf VB}\sim 50^\circ-60^\circ$ (see discussion below). The second important observation is that the density fluctuations seem to reinforce the anisotropy of the cascade rate w.r.t. the Alfv\'enic turbulence: The ratio $\cal R=\langle \vert \epsilon_C \vert\rangle/ \langle \vert \epsilon_I \vert \rangle$ (in blue) is close to $1$ for the Alfv\'enic cases,  but increases to $\sim 3$ for the magnetosonic ones at quasi-perpendicular angles. Numerical simulations of compressible MHD turbulence showed that fast magnetosonic turbulence is spatially isotropic while slow mode turbulence is anisotropic and has a spectrum ${k_\perp}^{-5/3}$ similarly to Alfv\'enic turbulence~\cite{Cho02}. This first observation that density fluctuations enhance the anisotropy of the cascade rate suggests that a slow-like (or mirror) mode turbulence dominates the compressible fluctuations analyzed here~\cite{Sahraoui2006,Hadid2015}.
\begin{figure}
\includegraphics[width=1\columnwidth]{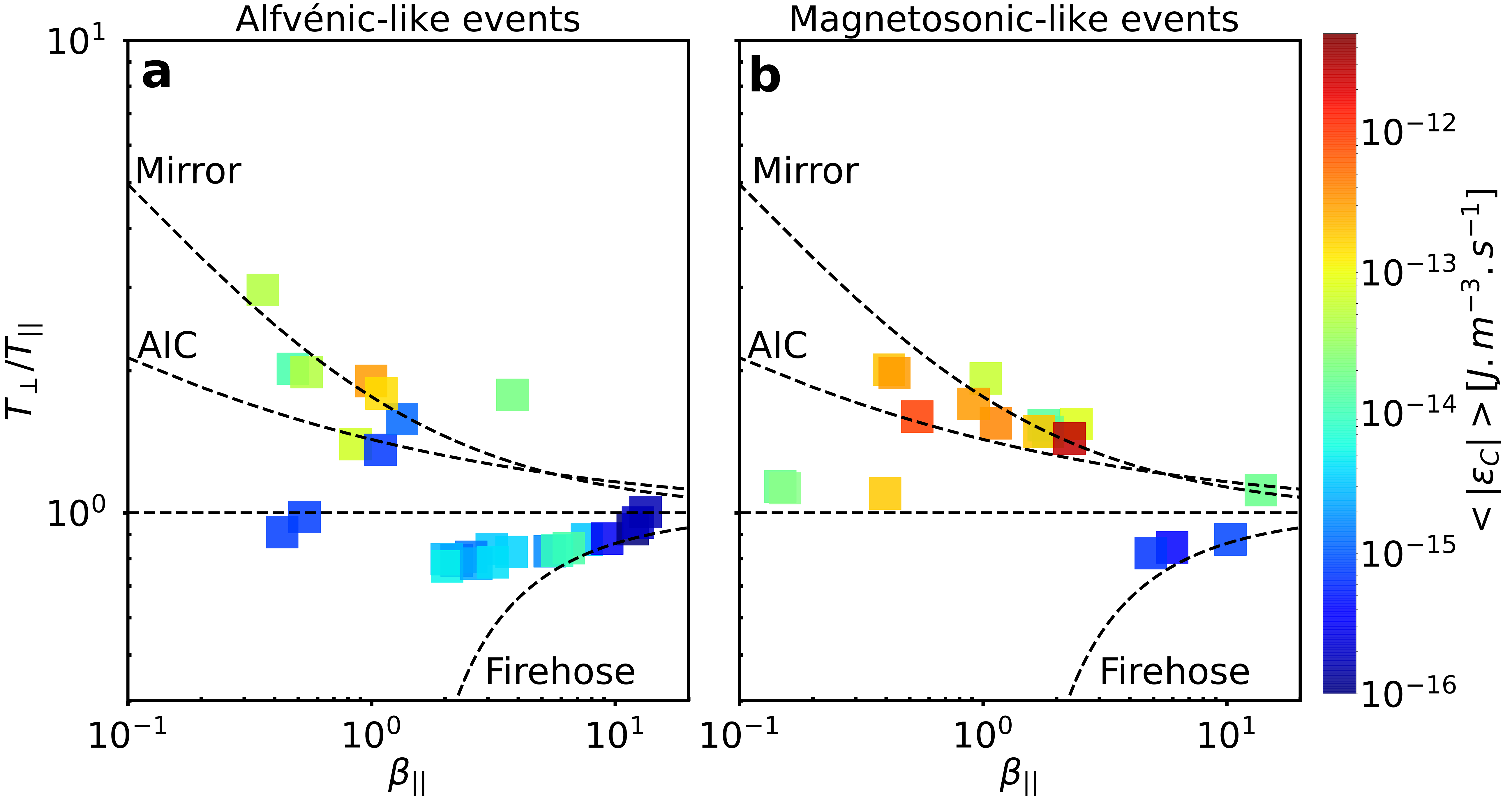}
\caption{Compressible energy cascade rate $\langle \vert \epsilon_{C}\vert \rangle $ averaged into bins of proton temperature anisotropy ($T_{\perp}/T_{\parallel}$) vs $\beta_{\parallel}$ for (a) the Alfv\'enic and (b) magnetosonic-like events. The dashed lines correspond to the mirror, the AIC, and firehose linear instabilities thresholds.}
\label{Temp_Anisotropy}  
\end{figure}
This result agrees with the analysis of the stability conditions of the plasma derived from the linear Maxwell-Vlasov theory. Figure \ref{Temp_Anisotropy}(a) shows that a large fraction of the Alfv\'enic-like cases that have the lowest values of the cascade rate correspond to a marginally stable plasma with $T_{\perp}/T_{\parallel} \sim 1$. This contrasts with the results from the magnetosonic turbulence (Figure \ref{Temp_Anisotropy}(b)) showing that most of the events have strong temperature anisotropy and lie near the mirror instability threshold, where the energy cascade rate is the highest. Considering that the maximum growth rate of the linear mirror instability occurs at oblique angles $\Theta_{\bf kB}$ (approximated here by the angle $\Theta_{\bf VB}$)~\cite{Pokhotelov2004}, the peak of the cascade rate and the total energy observed for $\Theta_{\bf VB}\sim 50^\circ-60^\circ$  in Fig.~\ref{Angle_Anisotropy} may be explained by energy injection into the background turbulent plasma through the mirror instability, which seems to enhance the dissipation rate. A similar relationship between incompressible cascade rate and kinetic plasma instabilities was found in the solar wind~\cite{Osman2013}, however, deeper understanding of the connection between these two features of plasma turbulence requires further theoretical investigation~\cite{Kunz2014}. Although the Taylor hypothesis (implicitly used in this work to interpret time lags $\tau$ as spatial increments) cannot generally be tested in single spacecraft data, the dominance of anisotropic slow (or mirror) like modes and the intrinsic anisotropic nature of the Alfv\'enic turbulence are arguments in favor of the validity of the Taylor hypothesis~\cite{Howes2014}. Indeed, k-filtering results (not shown here) obtained from four samples of Cluster data intervals to which the technique could be applied, support this conclusion (see, e.g., \cite{Sahraoui2006}).

{\it Conclusions ---} The energy cascade rate in MHD turbulence in a the compressible magnetosheath plasma was found to be at least two orders of magnitude higher than in the (nearly) incompressible solar wind. Empirical laws relating the cascade rate to the turbulent Mach number were obtained. Density fluctuations were shown to amplify magnitude and the spatial anisotropy of the cascade rate in comparison with incompressible Alfv\'enic turbulence. This result and the analysis of the plasma stability conditions in the plane $(T_\perp/T_\parallel, \beta_\parallel)$ indicate that the density fluctuations are carried by mirror (slow magnetosonic-like) mode driven by proton temperature anisotropy. These new fundamental features of compressible turbulence may have potential applications in the magnetostheath (e.g, turbulence-driven reconnection at the magnetopause \cite{Belmont2001,Karimabadi2014}) and in distant astrophysical plasmas. For instance, recently \citet{Zank2017} showed the importance of the compressible magnetosonic modes in forming the turbulent energy cascade in the Local ISM, the Heliosheath (also a bounded region, by the termination shock and the Heliopause) using in-situ Voyager 1 data, results similar to the ones reported here.

Helpful discussions with N. Andres, S. Banerjee and H. Breuillard are gratefully acknowledged. The fields and the particles data of CLUSTER spacecraft come from the CAA (ESA) and of THEMIS B/C spacecraft from the AMDA science analysis system provided by the Centre de Données de la Physique des Plasmas (IRAP, Université Paul Sabatier, Toulouse) supported by CNRS and CNES (CDPP, IRAP, France). The French participation in Cluster and THEMIS are supported by CNES.

\end{document}